\documentclass{appolb}
\usepackage{epsfig}

\begin{document}
\title{Kinetics vs hydrodynamics: \\ generalization of Landau/Cooper-Frye prescription for freeze-out%
\thanks{Presented at Workshop on Particle Correlations and Femtoscopy: WPCF-2008}%
}
\author{Yu.M. Sinyukov, S.V. Akkelin, Iu.A. Karpenko
\address{Bogolyubov Institute for Theoretical
Physics, Metrolohichna str. 14b, 03680 Kiev-143,  Ukraine}
\and
Y. Hama
\address{Instituto de F\'{\i}sica,
Universidade de S\~ao Paulo, S\~ao Paulo, SP, C.P. 66318,
05314-970, Brazil}
} \maketitle

\begin{abstract}
The problem of spectra formation in hydrodynamic approach to A+A
collisions is considered within the Boltzmann equations. It is
shown analytically and illustrated by numerical calculations that
the particle momentum spectra can be presented in the Cooper-Frye
form despite freeze-out is not sharp and has the finite temporal
width. The latter is equal to the inverse of the particle
collision rate at points $(t_{\sigma}({\bf r},p),{\bf r})$ of the
maximal emission at a fixed momentum $p$. The set of these points
forms the  hypersurfaces $t_\sigma({\bf r},p)$ which strongly
depend on the values of $p$ and typically do not enclose
completely the initially dense matter. This is an important
difference from the standard Cooper-Frye prescription (CFp), with
a common freeze-out hypersurface for all $p$, that affects
significantly the predicted spectra. Also, the well known problem
of CFp as for negative contributions to the spectra from
non-space-like parts of the freeze-out hypersurface is naturally
eliminated in this improved prescription.
\end{abstract}

\section{Introduction}

The Landau hydrodynamic approach \cite{Landau} for multi-hadron
production in hadronic/nuclear collisions appeared as a method
that is alternative to the S-matrix one: the latter transforms the
asymptotic hadronic states from $t= - \infty $ to $t=+ \infty $
while the former deals with the space-time evolution of thermal
matter, produced in the collisions. The evolution is described by
using the local energy-momentum conservation laws and
thermodynamic equation of state of the matter. Unlike the S-matrix
formalism, the initial state in hydrodynamic approach is
associated with some concrete finite time after collision when
created particles reach the locally equilibrated state and so, the
initial state can be described by the minimal set of parameters.
As well, because system formed is quite small, the picture of
continuous medium is destroyed also at likely small finite time.
Since the system expands fast, the latter is not constant in
configuration space but depends on the position ${\bf r}$ of the
fluid elements: $t_{\sigma}({\bf r})$. The set of these points
$(t_{\sigma}({\bf r}), {\bf r})$ in the Minkowski space forms,
therefore, the hypersurface $\sigma$ corresponding to the outer
boundary of the applicability of hydrodynamics. The sudden
freeze-out implies that the spectra are formed just on this
hyperurface where, formally, an ideal fluid transforms into an
ideal gas. The particle momentum spectra then can be expressed by
the well-known Cooper-Frye formula \cite{Cooper}:
\begin{eqnarray}
 p^0 n(p) =p^0\frac{d^3N}{d^3p}\approx
\mathrel{\mathop{\int }\limits_{\sigma}}
d\sigma_{\mu}p^{\mu}f_{\tt{l.eq.}}(x,p). \label{CFp}
\end{eqnarray}
The freeze-out hypersurface is typically associated with an
isotherm. According to Landau \cite{Landau} $T\approx
m_{\pi}\approx$ 140 MeV. In current analysis of A+A collisions the
corresponding temperature is in the region $T=90-150$ MeV
and is fixed, typically, from the best fit of the spectra.

Any isotherm contains usually non-space-like parts, which lead to
unphysical negative contributions to spectra for the particles
with momenta directed inward the system,  $p_{\mu}d\sigma_{\mu}<
0$. There is no common phenomenological prescription, based on
Heaviside step functions $\theta(p_{\mu}d\sigma_{\mu})$, which
allows one to eliminate the negative contributions to the momentum
spectra when $p^{\mu}d\sigma_{\mu}< 0$. The prescription, proposed
in \cite{Sin4}, eliminates the negative contributions in the way
which preserves the number of particles in the fluid element
crossing the freeze-out hypersurface. Therefore it takes into
account that at the final stage the system is the only holder of
emitted particles. Another prescription \cite{Bugaev} ignores the
particle number conservation considering decaying hadronic system
rather as a star - practically unlimited reservoir of emitted
photons/particles. The both prescriptions have a problem with
momentum-energy conservation laws at  freeze-out.

But the most serious problem is the obvious conflict of the CFp
with simple observation: to provide sudden transformation of the
liquid to ideal gas one needs to switch suddenly cross-sections
between particles from very big values to very small ones
\cite{AkkSin1} that cannot happen in reality. The reason of
particle liberation is another: it is the gradual change of the
ratio between the rate of system expansion and the rate of
collisions, so that particles should be emitted continuously. The
phenomenological model of continuous emission was proposed in Ref.
\cite{Hama2}; another approach is the so-called hybrid models
\cite{hybrid},  where kinetic evolution is matched with hydro one
at the hypersurface of hadronization. The shortcomings of these
models are described in detail in \cite{PRC}. But the feeling that
the Landau/Cooper-Frye sudden freeze-out does not describe the
real process of continuous particle emission is predominant now.

Nevertheless, in Ref. \cite{dual} it was advanced an idea  of
duality in hydro-kinetic approach to A+A collisions: though the
process of particle liberation, described by the emission
function, is non-equilibrium and gradual, the observable spectra
can yet be expressed by means of the Cooper-Frye prescription
based on locally equilibrium distribution function. The conclusive
step towards an analytical and numerical realization of this idea
is done in Ref. \cite{PRC} where hydro-kinetic approach is
developed: it overcomes all above mentioned problems by
considering the continuous dynamical freeze-out that is consistent
with Boltzmann equations and conservation laws. In what follows we
stick to the mainstream of arguments developed in this paper.

\section{Kinetics of freeze-out in Boltzmann approach}

Let us start from Boltzmann equation. It has the general form:
\begin{eqnarray}
\frac{p^{\mu }}{p^{0}}\frac{\partial f_i(x,p)}{\partial x^{\mu }}
= G_i(x,p)-L_i(x,p). \label{diff}
\end{eqnarray}

The expressions $G_i(x,p)$ and $L_i(x,p)= R_i(x,p)f_i(x,p)$ are
so-called (G)ain  and (L)oss terms for the particle of species
$i$. Typically, $R_i$ is a rate of collisions of $i$-th particle.
Below we will omit index $i$, the corresponding expression can be
related then, e.g., to pions.

The probability ${\cal P}_{t\rightarrow{t^{\prime}}}(x,p)$ for a
particle to reach the point $x^{\prime}=(t^{\prime},{\bf
r}^{\prime})$ starting from the point $x=(t,{\bf r})$ without
collisions is

\begin{equation}
{\cal P}_{t\rightarrow{t^{\prime}}}(x,p)=\exp\left(  -\int\limits_{t}^{t^{\prime}%
}d{\overline{t}}R({\overline{x}_{t}},p)\right)\label{probab},
\end{equation}
where $$\overline{x}_{t}=(\overline{t}, {\bf r}+\frac{{\bf
p}}{p^0}(\overline{t}-t)).$$ In terms of this probability, the
Boltzmann equation can be rewritten in the following integral form
\begin{eqnarray}
f(t,{\bf r},p)= f(t_0,{\bf r}-\frac{{\bf p}}{p^0}(t-t_0
), p){\cal P}_{t_0\rightarrow t}(t_0,{\bf r}-\frac{{\bf p}}{p^0}(t-t_0),p)  \nonumber\\
{+\int\limits_{t_0}^{t}%
}G(\tau,{\bf r}-\frac{{\bf p}}{p^0}(t-\tau),p){\cal
P}_{\tau\rightarrow t}(\tau,{\bf r} -\frac{{\bf
p}}{p^0}(t-\tau),p)d\tau . \label{integr}
\end{eqnarray}

Let us integrate the distribution (\ref{integr}) over the space
variables to represent the particle momentum density at large
enough time, $t\rightarrow\infty$, when particles in
 the system stop to interact. To simplify notation let us introduce
the escape probability for the particle with momentum $p$ in the
point $x=(t,{\bf r})$ to leave system without collisions: ${\cal
P}(t, {\bf r},p)\equiv {\cal P}_{t\rightarrow (\tau \rightarrow
\infty)}(t,{\bf r},p)$. Then the result can be presented in the
general form found in Ref. \cite{AkkSin1}:

\begin{eqnarray}
n(t\rightarrow \infty,p)\equiv n(p) = \int d^{3}r f(t_0,{\bf
r},p){\cal
P}(t_0,{\bf r},p)\nonumber \\
+ \int d^{3}r{\int\limits_{t_0}^{\infty}}
dt^{\prime}G(t^{\prime},{\bf r},p){\cal P}(t^{\prime},{\bf r},p).
\label{spectrum}
\end{eqnarray}

 The first term in Eq. (\ref{spectrum}) describes the
contribution to the momentum spectrum from particles that are
emitted from the very initial time, while the second one
describes the continuous emission with emission density
 $S(x,p)=G(t,{\bf r},p){\cal P}(t,{\bf r},p)$ from 4D volume
 delimited by the initial and final (where particles stop to interact) 3D hypersurfaces.

In what follows we will use the (generalized) relaxation time
approximation proposed in \cite{AkkSin1}, which is the basis of
the hydro-kinetic approach, described in detail in \cite{PRC}.
Namely, it was argued \cite{AkkSin1} that there is such a local
equilibrium distribution function
$f_{l.eq.}(T(x),u^{\nu}(x),\mu(x))$ that, in the region of not
very small densities where term $G\sim S$ gives noticeable
contribution to particle spectra, the function $f$ is
approximately equal to that one which would be obtained if all
functions in r.h.s. of Eq. (\ref{integr}) calculated by means of
that function $f_{l.eq.}$. The function $f_{l.eq.}$ is determined
from the local energy-momentum conservation laws based on the
non-equilibrium function $f$ in the way specified in \cite{PRC}.
Then, in accordance with this approach we use
\begin{equation}
R(x,p)\approx R_{l.eq.}(x,p), G\approx
R_{l.eq.}(x,p)f_{l.eq.}(x,p). \label{leq}
\end{equation}
The ``relaxation time" $\tau_{rel}=1/R_{l.eq.}$ grows with time in this method.

\section{Saddle point approximation for momentum spectra}

Let us generalize now the Landau/Cooper-Frye prescription (CFp) of
sudden freeze-out. For this aim we apply the saddle point method
to calculate the integral in the expression for spectra
(\ref{spectrum}) with account of (\ref{leq}). To simplify notation
we neglect the contribution to the spectra from hadrons which are
already free at the initial thermalization time $t_0 \sim 1$ fm/c
and thus omit the first term in (\ref{integr}).

To provide straightforward calculations leading  to the
Cooper-Frye form let us shift the spacial  variables, ${\bf
r}^{\prime}={\bf r}+\frac{{\bf p}}{p_{0}}(t_0-t^{\prime})$, in
(\ref{spectrum}) aiming to eliminate the variable $t^{\prime}$ in
the argument of the function $R$ which is the integrand in ${\cal
P}(t^{\prime},{\bf r},p)$. Then
\begin{eqnarray}
n(p)\approx
 \int d^{3}r^{\prime} \stackrel{\infty}{%
\mathrel{\mathop{\int }\limits_{t_{0}}}%
} dt^{\prime } f_{\tt{l. eq.}}(t^{\prime },{\bf r}^{\prime
}+\frac{{\bf p}}{p_{0}}(t^{\prime }-t_{0}),p)Q (
t^{\prime},\bf{r}^{\prime},p),
 \label{mom-sp2}
\end{eqnarray}
 where
\begin{eqnarray}
Q(t^{\prime},{\bf r}^{\prime},p)= R(t^{\prime },{\bf r}^{\prime }+\frac{{\bf p}}{p_{0}}(t^{\prime }-t_{0}),p)%
\exp \left\{ -\stackrel{\infty}{%
\mathrel{\mathop{\int }\limits_{t^{\prime }}}%
}R(s,{\bf r}^{\prime  }+\frac{{\bf p}}{p_{0}}(s-t_{0
}),p)ds\right\} \label{mom-sp3}
\end{eqnarray}
 Note that
\begin{eqnarray}
Q(t^{\prime},{\bf r}^{\prime},p) = \frac{d}{dt^{\prime }}
P(t^{\prime},\bf{r}^{\prime},p),
 \label{mom-spectr-1}
\end{eqnarray}
where $P(t^{\prime},{\bf r}^{\prime},p)$ is connected with the
escape probability ${\cal P}$:
\begin{eqnarray}
P(t^{\prime},{\bf r}^{\prime},p)= {\cal P}(t^{\prime},{\bf
r}^{\prime }+\frac{{\bf p}}{p_{0}}(t^{\prime}-t_{0 }),p).
 \label{mom-spectr-2}
\end{eqnarray}
Therefore
\begin{equation}
\stackrel{\infty}{%
\mathrel{\mathop{\int }\limits_{t_0}}}dt^{\prime}Q(t^{\prime},{\bf
r}^{\prime},p)=1-{\cal P}(t_0,{\bf{r}}^{\prime},p)\approx 1.
\label{norm}
\end{equation}

The saddle point $t_{\sigma }({\bf r},p)$ is defined by the
standard conditions:
\begin{eqnarray}
\frac{d Q ( t^{\prime},{\bf r}^{\prime},p)}{d
t^{\prime}}|_{t^{\prime }=t_{\sigma}^{\prime}}=0,
 \nonumber \\
\frac{d^{2} Q ( t^{\prime},{\bf r}^{\prime},p)}{d t^{\prime
 2}}|_{t^{\prime }=t_{\sigma}^{\prime}}<0. \label{max}
\end{eqnarray}
Then one can get from (\ref{mom-spectr-1}), (\ref{mom-spectr-2})
the condition of the maximum of emission:
\begin{eqnarray}
- \frac{p^{\mu}\partial_{\mu}R(t^{\prime},{\bf
r},p)}{R(t_{\sigma}^{\prime},{\bf r},p)}
|_{t^{\prime }=t_{\sigma}^{\prime}, {\bf r}={\bf
r}^{\prime}+\frac{{\bf p}}{p_{0}}(t_{\sigma}^{\prime}-t_{0})}=
p_{0}R(t_{\sigma}^{\prime},{\bf r}^{\prime} +\frac{{\bf
p}}{p_{0}}(t_{\sigma}^{\prime}-t_{0}),p)\,. \label{6-8}
\end{eqnarray}

If one neglects terms ${\bf p}^*\partial_{{\bf r}^*}R$ in l.h.s.
and supposes that in the rest frame (marked be asterisk) of the
fluid element  with four-velocity $u(x)$ the collision rate,
$R^*(x,p)= \frac{p_{0}R(x,p)}{p^{\mu}u_{\mu}}$, does not depend on
particle momentum: $R^*(x)\approx \langle v^* \sigma \rangle (x)
n^*(x)$ (here $n(x)$ is particle density, $\sigma$ is the particle
cross-section, $v$ is the relative velocity, $<...>$ means the
average over all momenta), then the conditions (\ref{6-8}) are
equivalent to the requirement that at the temporal point of
maximum of the emission function the rate of collisions is equal
to the rate of system expansion \cite{PRC}. This is the heuristic
freeze-out criterion for sudden freeze-out \cite{Bondorf}.
However, as we will demonstrate, the neglect of momentum
dependence leads to quite significant errors.

To pass to the Cooper-Frye representation we use the variables
which include the saddle point:
\begin{eqnarray}
{\bf r} = {\bf r}^{\prime }+\frac{{\bf p}}{p_{0}}(t^{\prime
}_{\sigma }({\bf r}^{\prime },p)-t_{0}). \label{var}
\end{eqnarray}
Then the expression for the spectrum takes the form:
\begin{eqnarray}
n(p) \approx
 \int d^{3}r \left |1 -
\frac{\mathbf{p}}{p_{0}}\frac{\partial t_{\sigma }}{\partial {\bf
r}}\right | \stackrel{\infty}{%
\mathrel{\mathop{\int }\limits_{t_{0}}}%
} dt^{\prime } S(t^{\prime},{\bf r},p),
 \label{sps}
\end{eqnarray}
where the emission density in saddle point representation is
($t_{\sigma}\equiv (t_{\sigma}({\bf r},p)$)
\begin{eqnarray}
S( t^{\prime},{\bf r},p)=f_{\tt{l.eq.}}(t^{\prime
},{\bf r}+\frac{{\bf p}}{p_{0}}(t^{\prime }-t_{\sigma}),p)\nonumber \\
\times R(t_{\sigma},{\bf r},p){\cal P}(t_{\sigma},{\bf
r},p)\exp(-(t^{\prime }-t_{\sigma})^{2}/2D^2(t_{\sigma},{\bf
r},p)). \label{spe}
\end{eqnarray}
According to Eq. (\ref{probab}) ${\cal P}(t_{\sigma},{\bf
r},p)=e^{-1}$, since the freeze-out zone is the region of the last
collision for the particle. Then the normalization condition for
$Q$ ($Q$ is presented by the bottom line in (\ref{spe})) allows
one to determine the temporal width of the emission at the point
$(t_{\sigma }({\bf r},p),{\bf r},p)$:
\begin{eqnarray}
D(t_{\sigma},{\bf
r},p)=\frac{e}{\sqrt{2\pi}}\frac{1}{R(t_{\sigma},{\bf
r},p)}\approx \tau_{\tt{rel}}(t_{\sigma},{\bf r},p). \label{D}
\end{eqnarray}

Therefore if the temporal homogeneity length $\lambda(t,{\bf
r},p)$ of the distribution function $f_{\tt{l.eq.}}$ near the
4-point $(t_{\sigma}({\bf r},p),{\bf r})$ is much larger than the
width of the emission zone, $\lambda(t_{\sigma},{\bf r},p)\gg
\tau_{\tt{rel}}(t_{\sigma},{\bf r},p)$, then one can approximate
$f_{\tt{l.eq.}}(t^{\prime },{\bf r}+\frac{{\bf
p}}{p_{0}}(t^{\prime }-t_{\sigma}),p)$ by
$f_{\tt{l.eq.}}(t_{\sigma },{\bf r},p)$ in Eq. (\ref{sps}) and
perform integration over $t'$  accounting for normalizing
condition (\ref{norm}). As a result we get from (\ref{sps}) and
(\ref{spe}) the momentum spectrum in a form similar to the
Cooper-Frye one (\ref{CFp}):
\begin{eqnarray}
 p^0 n(p) =p^0\frac{d^3N}{d^3p}\approx
\mathrel{\mathop{\int }\limits_{\sigma(p)}}
d\sigma_{\mu}p^{\mu}f_{\tt{l.eq.}}(x,p). \label{mom-sp14}
\end{eqnarray}

It is worthy to note that the representation of the spectrum through
emission function (\ref{spectrum}) is the result of the
integration of the total non-equi\-librium distribution function
$f(x,p)$, Eqs. (\ref{integr}), (\ref{leq}), over the asymptotical
hypersurface in time, while the approximate representation of the
 spectrum, Eq. (\ref{mom-sp14}), uses only the local equilibrium
 part $f_{\tt{l.eq.}}$ of the total function $f(x,p)$ at the set
  of  points of maximal  emission - at hypersurface
$(t_{\sigma}({\bf r},p),{\bf r})$.

\section{Generalized Cooper-Frye prescription}

Now let us summarize the conditions when the Landau/Cooper-Frye form
for sudden freeze-out can be used. They are the following:
\smallskip

i) For each momentum ${\bf p}$, there is a region of ${\bf r}$
where the emission function as well as the function $Q$, Eq.
(\ref{mom-sp3}), have a clear maximum.  The temporal width  of the
emission $D$, defined by Eq. (\ref{spe}), which is found to be
equal to the relaxation time (inverse of collision rate), should
be smaller than the corresponding temporal homogeneity length of
the distribution function: $\lambda(t_{\sigma},{\bf r},p) \gg
D(t_{\sigma},{\bf r},p)\simeq\tau_{\tt{rel}}(t_{\sigma},{\bf
r},p)$.
\smallskip

ii) The contribution to the spectrum from the residual region of
${\bf r}$, where the saddle point method (Gaussian approximation
(\ref{spe}) and/or condition $\tau_{\tt{rel}} \ll \lambda$) is
violated, does not affect essentially the particle momentum
density.
\smallskip

If these conditions are satisfied, then the momentum spectra can
be presented in the Cooper-Frye form {\it despite the fact that
actually it is not sudden freeze-out and the decoupling region has
a finite temporal width} $\tau_{rel}(t_{\sigma},{\bf r},p) $.

The analytical results as for the temporal width of the spectra
agree remarkably with the numerical calculations of pion emission
function within hydro-kinetic model (HKM) \cite{PRC}. For example,
near the point of maximum, $\tau = 16.5$ fm/c, $r= 0, p_{T}= 0.2$
GeV, the ``experimental" temporal width $D_{HKM}$ obtained by
numerical solution of the complete hydro-kinetic equations is
$D_{HKM}\approx 4.95 $ fm (see Fig. 1, left). Our theoretical
estimate is $D=\frac{e}{\sqrt{2\pi}R}\approx 5.00$ fm, since the
rate of collisions in this phase-space point is
$R(\tau_{\sigma}({\bf r},p)= 16.5$ fm/c, ${\bf r}=0, p_T $= 0.2
GeV, $p_L =0)\approx 0.217$ c/fm.

It is worthy to emphasize that such a generalized Cooper-Frye
representation is related to {\it freeze-out hypersurfaces that
depend on the momentum} ${\bf p}$ and {\it typically do not
enclose the initially dense matter}. In Fig. 1, one can see the
structure of the emission domains for different $p_T$ in HKM
\cite{PRC} for initially (at $\tau$=1 fm/c) Gaussian energy
density profile with $\epsilon_{max}$= 6 GeV/fm$^3$. The maximal
emission regions for different $p_T$ are crossed by isotherms with
different temperatures: 80 MeV for low momenta and 135 MeV for
high ones. This is completely reflected in the concave structure
of the transverse momentum spectrum as one can see in Fig. 2.

If a part of the hypersurface $t_\sigma({\bf r},p)$ is
non-space-like and corresponds to the maximum of the emission of
particles with momentum ${\bf p}$, directed outward the system,
the same part of the hypersurface cannot correspond to the maximal
emission for particles with momentum directed inward the system.
It is clear that the emission function at these points is close to
zero for such particles. Even formally, in the Gaussian
approximation (\ref{spe}) for $Q$, validated in
 the region of its maximal value, the integral
$\stackrel{\infty}{%
\mathrel{\mathop{\int }\limits_{t_{\sigma}}}%
}dsR(s,{\bf r}+\frac{{\bf p}}{p^0}(s-t_{\sigma}({\bf r},p),p)\gg
1$, if particle world line crosses almost the whole system. The
latter results in $Q\rightarrow0$ and, therefore, completely
destroys the saddle-point  approximation (\ref{max}) for  $Q$ and
then the Cooper-Frye form (\ref{mom-sp14}) for spectra.
Recall that if a
particle crosses some non-space-like part of the hypersurface
$\sigma$ moving inward the system, this corresponds to the condition
$p^{\mu}d\sigma_{\mu}< 0$ \cite{Sin4}. Hence the value
$p^{\mu}d\sigma_{\mu}(p)$ in the generalized Cooper-Frye formula
(\ref{mom-sp14}) should be {\it always} positive:
$p^{\mu}d\sigma_{\mu}(p)>0$ across the hypersurface where fairly
sharp maximum of emission of particles with momentum ${\bf p}$ is
situated;  and so requirement $p^{\mu}d\sigma_{\mu}(p)>0$ is a
necessary condition for $t_{\sigma}({\bf r},p)$ to be a true
hypersurface of the maximal emission. It means that hypersurfaces
of maximal emission for a given momentum ${\bf p}$ may be open in
the space-time, not enclosing the high-density matter at the
initial time $t_0$, and different for different ${\bf p}$. All
this is illustrated in Fig. 3, where the structure of particle
emission domain is shown for two groups of particles. In the first
one, the momentum is directed as the radius vector to the point of
particle localization (they move outward the system), in the
second one - in opposite direction (they move inward). The points
of maximum for different $p_T$, where Cooper-Frye form can be applied,
do not overlap. The calculations have been done in
HKM$\,$\cite{PRC}.

Therefore,
there are no negative contributions to the particle momentum
density from non-space-like sectors of the freeze-out
hypersurface, that is a well known shortcoming of the Cooper-Frye
prescription \cite{Sin4,Bugaev}; the negative contributions could
appear only as a result of utilization of improper freeze-out
hypersurface that roughly ignores its momentum dependence and so
is common for all ${\bf p}$. If, anyhow, such a common
hypersurface will be used, e.g. as the hypersurface of the maximal
particle number emission (integrated over ${\bf p}$), there is no
possibility to justify the approximate expression for momentum
spectra similar to Eq. (\ref{mom-sp14}).

\section{Conclusions}

Our analysis and numerical calculations show that the widely used
phenomenological Landau/ Cooper-Frye prescription for calculation
of pion (or other particle) spectrum is too rough if the
freeze-out hypersurface is considered as common for all momenta of
pions. The Cooper-Frye formula, however, could be applied in
generalized form accounting for direct momentum dependence of the
freeze-out hypersurface $\sigma (p) $; the latter corresponds to
the maximum of emission function $S(t_{\sigma}({\bf r},p),{\bf
r},p)$ at fixed momentum ${\bf p}$ in an appropriate region of
${\bf r}$.
 If such a hypersurface $\sigma (p) $ is found, the condition of
applicability of the Cooper-Frye formula {\it for given} ${\bf p}$
is that the width of the maximum, which in the simple cases -
e.g., for one component system or at domination of elastic
scatterings - is just the relaxation time (inverse of collision
rate), should be smaller than the corresponding temporal
homogeneity length of the distribution function.

\section*{Acknowledgments}

This work was supported in part by FAPESP (Brazil), under the
contract numbers 2004/10619-9 and 2008/55658-2; the Fundamental
Research State Fund of Ukraine, Agreement No. F25/239-2008; the
Bilateral award DLR (Germany) - MESU (Ukraine) for the UKR 06/008
Project, Agreement No. M/26-2008; and the Program ``Fundamental
Properties of Physical Systems under Extreme Conditions"  of the
Bureau of the Section of Physics and Astronomy of NAS of Ukraine.

\newpage
\begin{figure}\label{fig1}
 \vspace{-0.3cm}
 \hspace{-2cm}
 \includegraphics[scale=0.4]{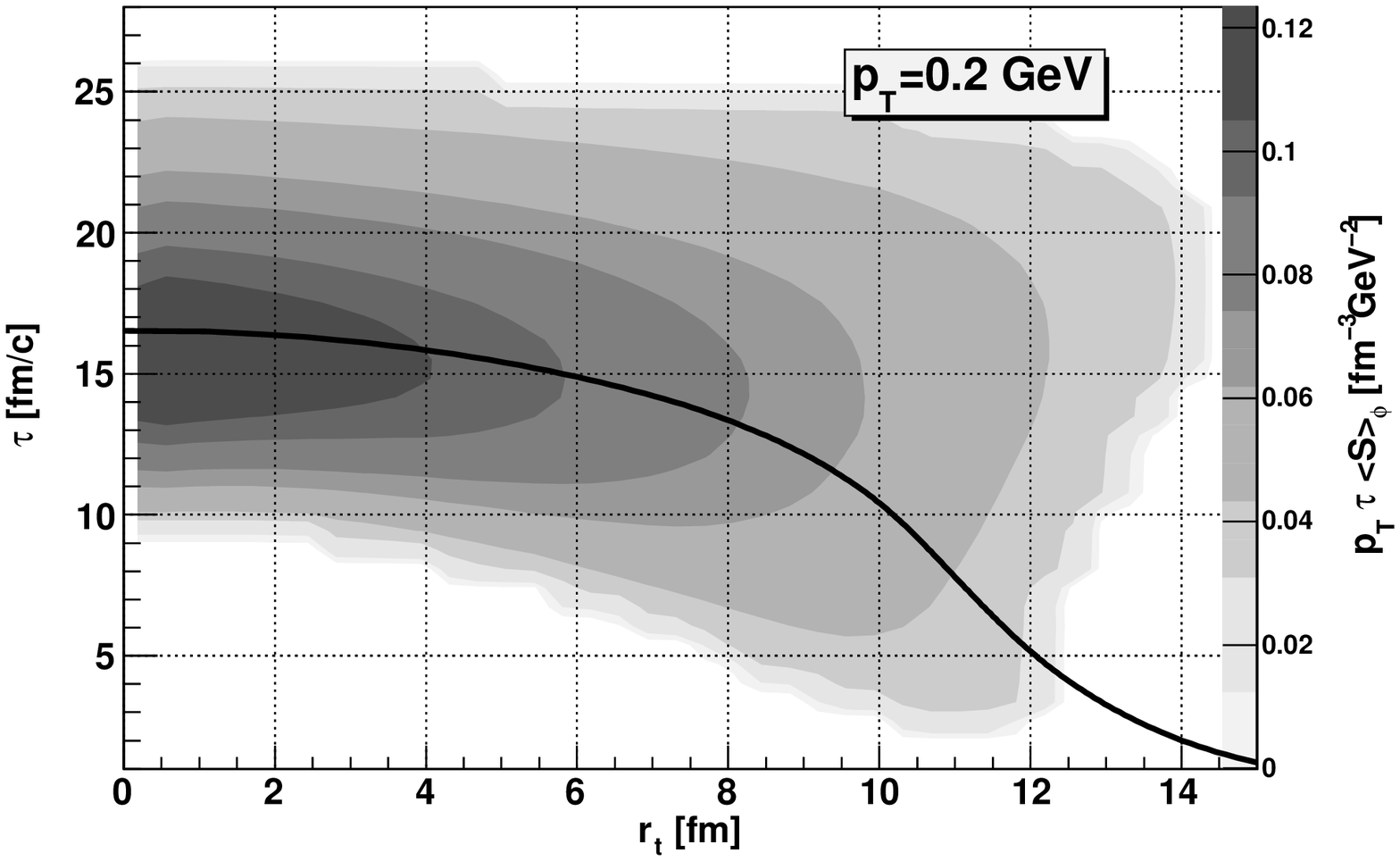}
 \includegraphics[scale=0.4]{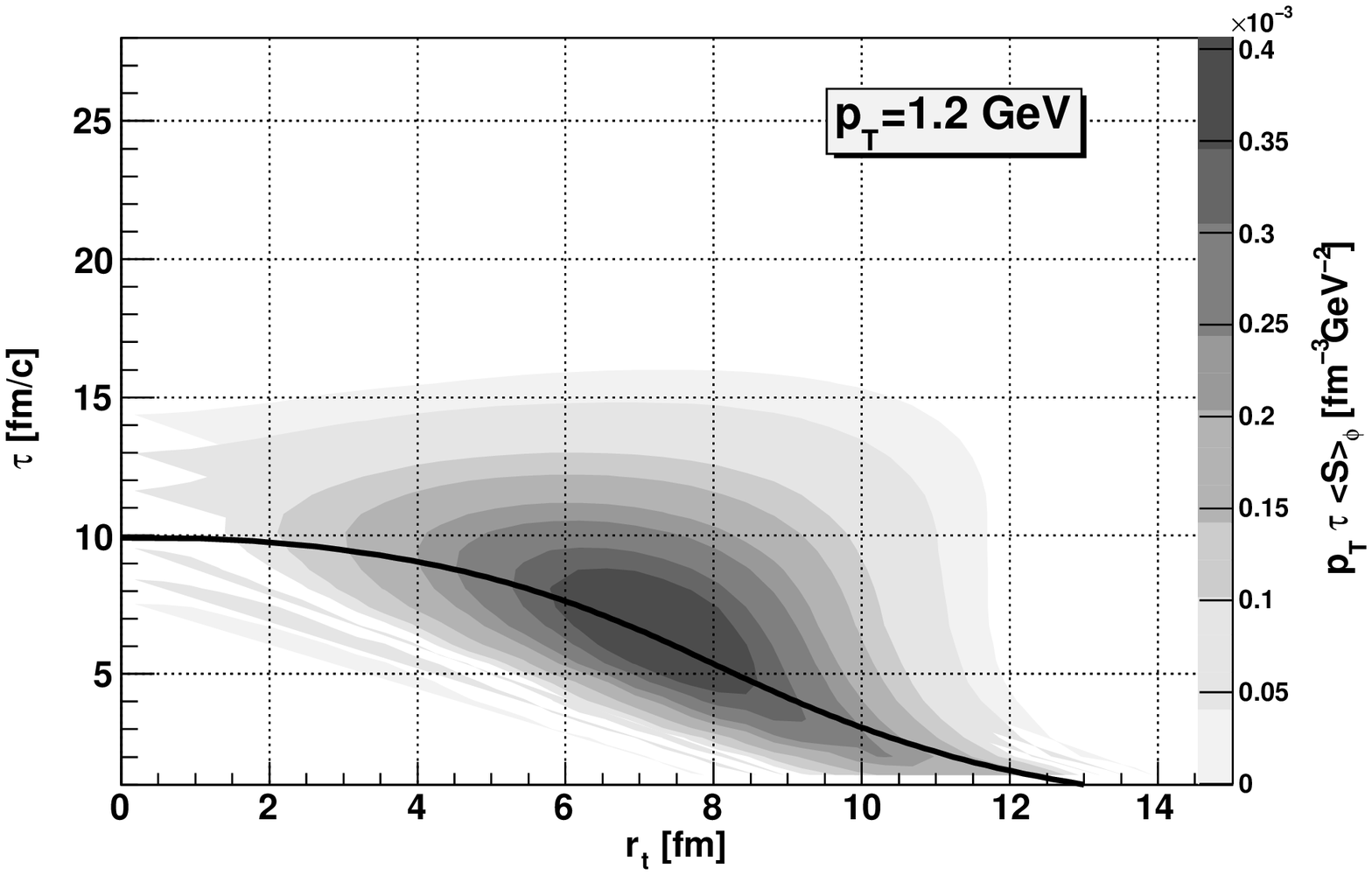}\hspace{-2cm}
 \vspace{-0.7cm}
 \caption{The pion emission function for different $p_T$ in
 hydro-kinetic model (HKM) \cite{PRC}. The isotherms of 80
 MeV (left) and 135 MeV (right) are superimposed.}
\end{figure}

\begin{figure}\label{fig2}
 \vspace{-0.4cm}
 \hspace{2cm}
 \includegraphics[scale=0.4]{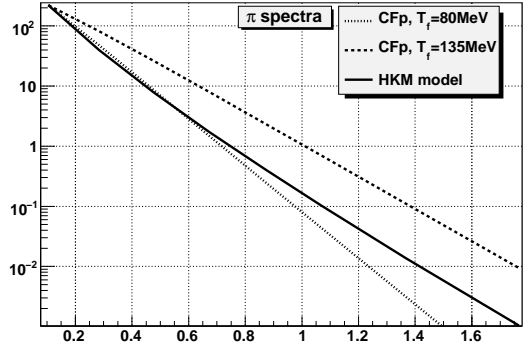}
 \vspace{-0.2cm}
 \caption{Transverse momentum spectrum of $\pi^{-}$ in HKM,
 compared with the sudden freeze-out ones at temperatures of 80 and 160 MeV with arbitrary normalization.}
\end{figure}

\begin{figure}\label{fig3}
 \vspace{-0.3cm}
 \hspace{-2cm}
 \includegraphics[scale=0.4]{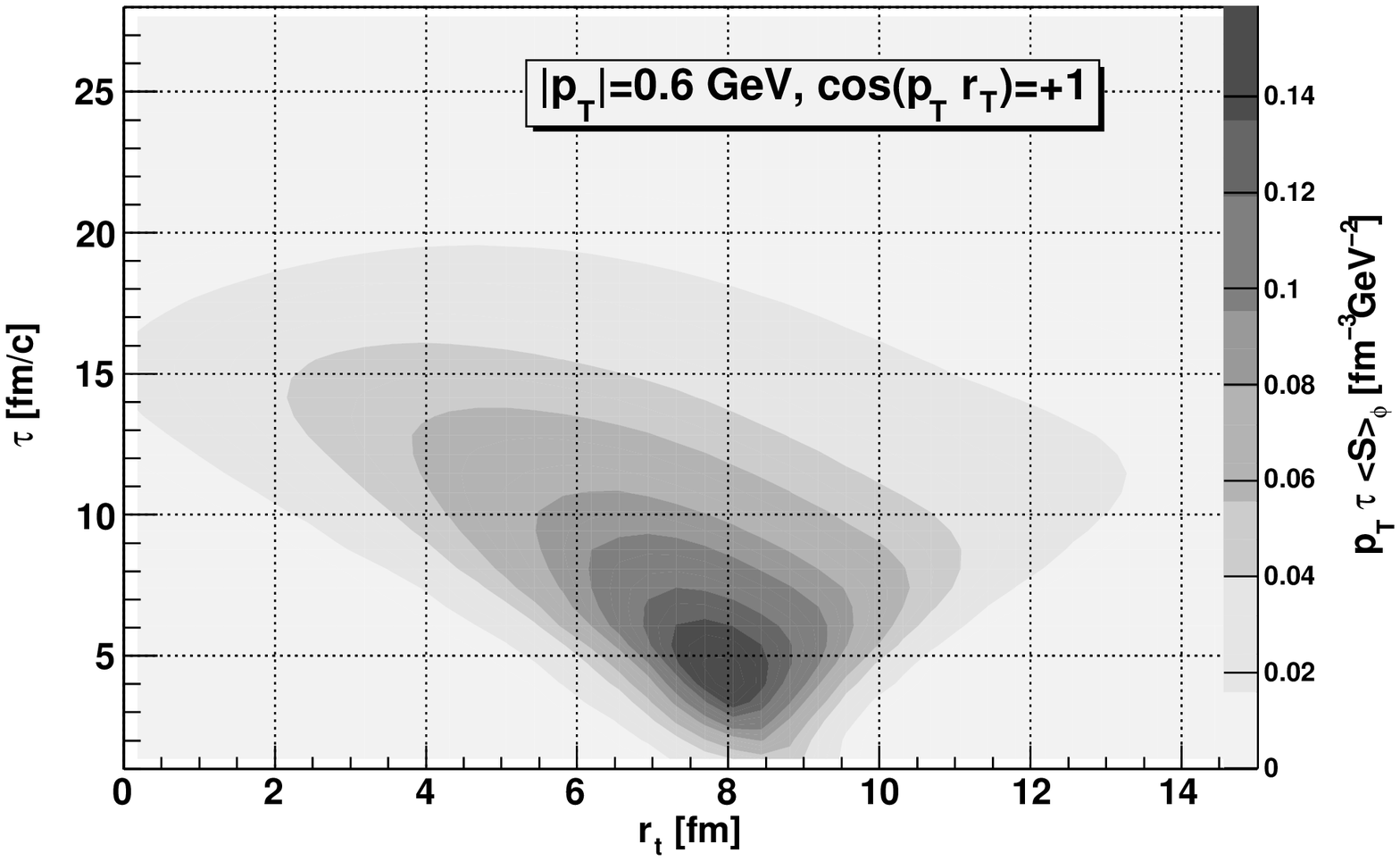}
 \includegraphics[scale=0.4]{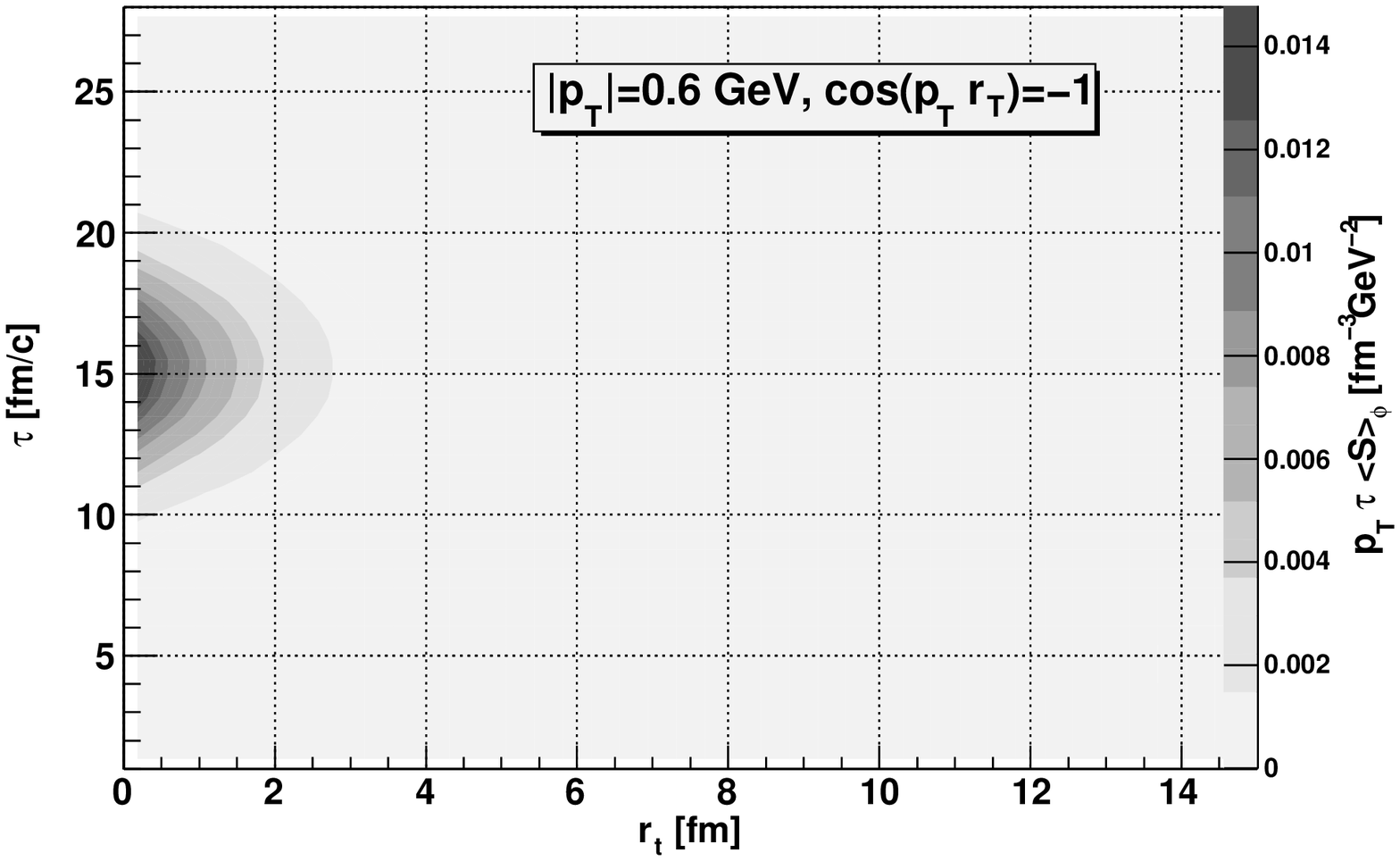}\hspace{-2cm}
 \vspace{-0.7cm}
 \caption{The emission function in HKM for particles with momentum directed along the radius  vector at
the emission points (left) and for those ones in the
 opposite direction to the radius vector (right).}
\end{figure}

\end{document}